\documentclass[aps,pre,twocolumn,showpacs,amssymb]{revtex4}

\usepackage[dvips]{graphicx}\usepackage{dcolumn}
\usepackage{bm}
\def\en{\end{equation}}

\def\bea{\begin{eqnarray}}
\def\ena{\end{eqnarray}}

\usepackage{color}

\begin{document}


\title{Estimation of the bending rigidity and spontaneous curvature of fluid membranes in simulations}



\author{Hayato Shiba}
\author{Hiroshi Noguchi}
\email{noguchi@issp.u-tokyo.ac.jp}
\affiliation{Institute for Solid State Physics, University of Tokyo, Chiba 277-8581, Japan}


\date{\today}

\begin{abstract}
Several numerical methods for measuring the bending rigidity and the
spontaneous curvature of fluid membranes are studied using 
two types of meshless membrane models.
The  bending rigidity is estimated from
the thermal undulations of planar and tubular membranes
and the axial force of tubular membranes.
We found a large  dependence
of its estimate value from the thermal undulation analysis
on the upper-cutoff frequency $q_{\rm {cut}}$ of the least squares fit.
The inverse power-spectrum fit with an extrapolation to $q_{\rm {cut}} \to 0$
yields the smallest estimation error among the investigated methods.
The spontaneous curvature is estimated from
the  axial force of tubular membranes and the average curvature of bent membrane strips.
The results of these methods show good agreement with each other.
\end{abstract}


\maketitle

\section{Introduction}

When amphiphilic molecules
are dissolved into aqueous environments,
these molecules self-assemble into
several types of characteristic structures
such as
spherical or cylindrical micelles, bilayers, and 
inverted hexagonal structures
\cite{SafranBook,GompperSchick,LipowskySackmann,rand90}.
Among them, 
bilayer membrane is the basic structure
of cells and organella.
In living cells, biomembranes are not only static walls that separate components
but also dynamical objects playing functions
such as the vesicle transport of proteins via membrane fusion and fission.

On a micrometer scale, 
the lipid membranes can be considered as a continuum surface,
where the membrane thickness can be neglected.
The curvature free energy of a curved membrane
is  given by \cite{Helfrich}
\begin{equation}
\mathcal{F} = \int \left[ \frac{\kappa}{2} (C_1 +
C_2- C_0)^2 + \overline{\kappa} C_1C_2\right] dA,
\label{eq:Helfrich}
\end{equation}
where $C_1$ and $C_2$ are the principal curvatures 
at each position of the membrane. 
The coefficients $\kappa$ and $\bar{\kappa}$
are the bending rigidity and saddle-splay modulus, respectively.
The spontaneous curvature $C_0$ vanishes when lipids symmetrically 
distribute in both leaflets of the bilayer.
The last term in Eq. (\ref{eq:Helfrich}) is constant for
a fixed topology (Gauss-Bonnet theorem). 
The bending rigidity and spontaneous curvature
 are basic quantities to understand the membrane properties.
In this paper, we study the numerical measurement methods of 
the bending  rigidity $\kappa$ and the spontaneous curvature $C_0$
in simulations.

Several methods have been used to measure the bending rigidity $\kappa$
in experiments and simulations.
They are classified to two groups:
(i) One utilizes the thermal 
fluctuations of the undulation modes of the 
membrane surface. In experiments,
the surface fluctuations are measured by light microscopy 
 with vesicles, cells,  
etc \cite{Schneider1984,Fricke,Faucon,Duwe,Meleard}. 
Theoretically, the fluctuation spectrum is 
derived by the perturbations from  planar \cite{SafranBook,sack94,Goetz1999}, 
spherical \cite{helf86,Safran1987,seif97,gg:gomp04c}, and 
cylindrical \cite{Ouyang1989,Fournier2007,Fournier2009} membranes.
In simulations, the  fluctuation spectrum of planar membranes
is widely used to measure $\kappa$ 
\cite{Goetz1999,lind00,Marrink2001,Farago2003,stec04,Briels2005,nogu11}.
The fluctuations of tubular membranes have not been simulated as yet,
whereas those of quasi-spherical vesicles are calculated
in Refs. \cite{gg:gomp04c,Drouffe,nogu01b}.
(ii) The other utilizes force measurements.
A tubular (tether) membrane is formed from a liposome
by a mechanical force (induced by optical tweezers, etc.).
The bending rigidity can be measured using
the force strength and the surface tension of the vesicle
\cite{Bo,Evans1990,Dai,Evans1996,Cuvelier}. 
The stability and the shapes of tubular membranes
 have been intensively studied in theories
\cite{Ouyang1989,Fournier2007,Fournier2009,Svetina1992,umed98,Bukman,Powers,Prost2002}.
In simulations, Harmandaris {\it et al.} \cite{Harmandaris} 
measured  $\kappa$ from the axial force and radius of tubular membranes.
Recently, $\kappa$ was also measured from the surface tension of
the buckled membranes in simulations \cite{nogu11a}.

In living cells, biomembranes have asymmetric lipid distribution
in two leaflets \cite{kamp79}. 
Such asymmetry of the membranes yields a non-zero spontaneous curvature $C_0$. 
For a closed membrane {\it i.e.} vesicle, 
a low flip-flop rate between the leaflets 
can result in an effective spontaneous curvature \cite{seif97}.
The vesicle morphology is varied with $C_0$ \cite{harb77,juli96,das09}. 
Recent experiments show that the spontaneous curvature 
$C_0$ is also induced by grafting polymers, by absorption of protein onto the membrane
surface, or by other means \cite{Baumgart2010,phil09,zimm06,akiy03,four09a}. 
Since many proteins were found to control membrane curvatures,
much attention has been paid to the effects of the spontaneous curvature $C_0$.
To simulate the effects of $C_0$,
it is important to establish numerical methods to measure $C_0$.
In previous studies, the spontaneous curvature
is estimated from the comparison of membrane shapes with 
the results of the continuum theory \cite{harb77,mark06}.
In this paper, we propose two direct methods for measuring the spontaneous curvature.

Many types of membrane models have been developed for simulations 
from the atomistic scale to a large micrometer scale 
(see review articles \cite{muel06,vent06,klie08,marr09,NogRev}).
Among them, 
particle-based meshless membrane models
\cite{Drouffe,NogRev,Nog06PRE,Nog06JCP,Popolo,kohy09,liu09,fuch09,2010Yuan1}
are suitable for studying the large-scale membrane dynamics 
including  topological changes such as membrane rupture, fusion, and fission.
In these meshless models,
a membrane particle does not represent a lipid molecule; it represents a membrane 
patch consisting of many molecules. 
The membrane particles self-assemble to form vesicles and planar
 membranes owing to their attractive interactions.

We employ two types of meshless models in this paper:
the mls membrane model \cite{Nog06PRE,Nog06JCP} and a new spin membrane model.
In contrast to the mls model, the spin model allows a finite spontaneous curvature 
$C_0$ similar to Yuan's meshless model \cite{2010Yuan1}.
In our meshelss models, the bending rigidity $\kappa$ and the line tension $\Gamma$ of the 
membrane edge can be varied separately for wide ranges of the fluid phase.

In Sec. \ref{sec:method}, the membrane models and the simulation methods are described.
The measurements of the bending rigidity $\kappa$ from the undulation 
mode analysis of planar membranes is explained in Sec. \ref{sec:flat}.
The stretching force measurement 
and thermal undulations of tubular membranes are described
in Sec. \ref{sec:nocv}.
The measurement of the spontaneous curvature $C_0$
from the force measurement of tubular membranes
and the average curvature of bent membrane strips is explained in Sec. \ref{sec:spcv}.
A summary is provided in Sec. \ref{sec:summary}.

\section{Simulation Model and Method} \label{sec:method}

We employ 
two types of meshless membrane models.
They use different curvature potentials.
(i) The meshless mls membrane model \cite{Nog06PRE,Nog06JCP}:
Membrane particles possess only translational degrees of freedom,
and form quasi-two-dimensional structures
stabilized by a mutibody potential
based on moving least-squares (mls) method \cite{Nog06PRE,Nog06JCP}.
(ii) The meshless spin membrane model:
Particles also possess orientational degrees of freedom
and interact with potentials similar to those described in Ref. \cite{nogu11}.
In both the models, 
the bending rigidity $\kappa$ and the line tension $\Gamma$ of the 
membrane edge are controlled by changing 
the parameters of particle interactions (see Appendix).  
Further, the spontaneous curvature $C_0$ can be varied in the spin model.

\subsection{Meshless mls membrane model}

Since the model is explained in detail in 
Refs. \cite{Nog06PRE,Nog06JCP}, the model is outlined
 only briefly in this section. 
A membrane consists of $N$ particles, which possess
no internal degrees of freedom. The particles interact 
with each other via a potential
\begin{equation}
\frac{U}{k_{\rm B}T} = \varepsilon (U_{\rm rep} + U_{\rm att} ) + U_\alpha,
\end{equation}
where  $k_{\rm B}T$ is the thermal energy.
The potential $U$ consists of a repulsive soft-core potential $U_{\rm {rep}}$, 
 an attractive potential $U_{\rm {att}}$, with a coefficient $\varepsilon$,
 and a curvature potential $U_{\alpha}$.
In a quasi-two-dimensional membrane surface, the particles 
interact via the potentials $U_{\textrm{\scriptsize{rep}}}$ and $U_{\textrm{\scriptsize{att}}}$. 
The repulsive excluded interaction potential of a diameter $\sigma$ is given by 
\begin{equation}
U_{\textrm{\scriptsize{rep}}} = \sum_{i<j} \exp ( -20 (r_{ij} /\sigma -1 ) + B ) f_{\rm {cut}} (r_{ij}/\sigma ),
\label{eq:urep}
\end{equation}
where $B=0.126$, and $r_{ij}$ is the distance between
particles $i$ and $j$. 
The diameter $\sigma$ is employed as the length unit 
to display the simulation results throughout this paper. 
A $C^\infty$-cutoff function 
\begin{equation}
f_{\textrm{\scriptsize{cut}}} (s) = \left\{
\begin{array}{ll}{}
\exp \{ a (1+ \frac{1}{(|s|/s_{\rm {cut}} )^n -1 }) \} & (s<s_{\rm {cut}} ) \\
0 & (s \ge s_{\rm {cut}} ) 
\end{array}
\right.  \label{eq:fcut}
\end{equation}
is employed.
For Eq. (\ref{eq:urep}), the values
$n=12, a=1,$ and $s_{\rm {cut}} = 1.2$ are used. 

A solvent-free membrane model requires an attractive interaction
mimicking the ``hydrophobic'' repulsion between hydrocarbon
chains of lipid or surfactant molecules and aqueous solvent.
We employ a potential 
\begin{equation}
U_{\textrm{\scriptsize{att}}} = \sum_{i} 0.25 \ln [1+ \exp \{ -4 (\rho_i -\rho^* ) \} ] -C,
\end{equation}
which is a function of the local density of the particles $\rho_i$, which are defined by
\begin{equation}
\rho_i = \sum_{j\neq i} f_{\textrm{\scriptsize{cut}}} (r_{ij}/\sigma).
\end{equation}
For this attractive interaction, 
the values 
$s_{\textrm{\scriptsize{half}}} = 1.8$
(at which $f_{\textrm{\scriptsize{cut}}}(s_{\textrm{\scriptsize{half}}} ) =0.5$),
$s_{\textrm{\scriptsize{cut}}} = s_{\textrm{\scriptsize{half}}} +0.3$, and $n=12$ are used.
The factor $a$ in $f_{\textrm{\scriptsize{cut}}} (s)$ is given by
$a = \ln (2) \{ 
( s_{ \textrm{\scriptsize{cut}}} / s_{\textrm{\scriptsize{half}}} )^n -1 \} \simeq 3.715$. 

The constant $C=0.25\ln \{ 1+ \exp (4\rho^*) \} \simeq \rho^*$
is set to achieve $U_{\textrm{\scriptsize{att}}}=0$ at $\rho_i=0$.
For $\rho_i <\rho^*$, the potential is approximately 
$U_{\textrm{\scriptsize{att}}} \simeq -\rho_i$, and therefore, it
acts as a pair potential with $U_{\textrm{\scriptsize{att}}} 
\simeq -\sum_{i<j} 2f_{\textrm{\scriptsize{cut}}} (r_{ij}/\sigma )$. 
For $\rho_i > \rho^*$, this function saturates to the constant $-C$. 
Thus, it is a pairwise potential with a smooth cutoff at the density 
$\rho_i=\rho^*$. 
We set $\rho^* =6$ in this paper to simulate
a fluid membrane.

In addition to the potentials $U_{\textrm{\scriptsize{rep}}}$ and 
$U_{\textrm{\scriptsize{att}}}$, we add a curvature potential 
\begin{equation}
U_\alpha = k_\alpha \sum_i \alpha_{\textrm{\scriptsize{pl}}} (\bm{r}_i),
\label{eq:ualpha} 
\end{equation}
where the shape parameter ``aplanarity'' is defined by
\begin{equation}
\alpha_{\textrm{\scriptsize{pl}}} = \frac{9D_{\rm {w}}}{T_{\rm {w}}M_{\rm {w}}}
= \frac{9\lambda_1\lambda_2\lambda_3}{ (\lambda_1+\lambda_2+\lambda_3)
(\lambda_1\lambda_2 +\lambda_2\lambda_3 + \lambda_3\lambda_1) }.
\end{equation}
The aplanarity $\alpha_{\textrm{\scriptsize{pl}}}$ represents 
the degree of deviation from the planar shape,
and $\lambda_1 \le \lambda_2 \le \lambda_3$ 
are the three eigenvalues of
the weighted gyration tensor for the $i$th particle given by
$a_{\alpha\beta} = \sum_j (\alpha_j -\alpha_{\rm G})(\beta_j-\beta_{\rm G}) w_{\textrm{\scriptsize{cv}}}(r_{ij})$,
where $\alpha,\beta \in \{x,y,z\}$
and the local center of mass ${\bf r}_{\rm G}=\sum_j {\bf r}_{j}w_{\rm {cv}}(r_{i,j})/\sum_j w_{\rm {cv}}(r_{i,j})$.
The aplanarity is calculated from three rotational invariants of the gyration tensor:
the determinant $D_{\rm {w}}$, trace $T_{\rm {w}}$,
and the sum of its three minors, 
$M_{\rm {w}}= a_{xx}a_{yy}+a_{yy}a_{zz}+a_{zz}a_{xx}
 -a_{xy}^2-a_{yz}^2-a_{zx}^2$.
This aplanarity $\alpha_{\textrm{\scriptsize{pl}}}$ takes 
a value in the interval $[0,1]$ and is proportional to $\lambda_1$
for $\lambda_1 \ll \lambda_2, \lambda_3$. 

A Gaussian function with a $C^{\infty}$ cutoff~\cite{Nog06PRE} 
is employed for calculation of the weight of the gyration tensor,
\begin{equation}
w_{\textrm{\scriptsize{cv}}}(r_{ij}) = 
\left\{
\begin{array}{ll}
\exp \Big( \frac{(r_{ij}/r_{\textrm{\scriptsize{ga}}} )^2}{(r_{ij}/r_{\textrm{\scriptsize{cc}}})^n -1} \Big) & (r_{ij}<r_{\textrm{\scriptsize{cc}}}) \\
0 & (r_{ij} \ge r_{\textrm{\scriptsize{cc}}})
\end{array}
\right. 
\end{equation}
which is smoothly cut off at $r_{ij} = r_{\textrm{\scriptsize{cc}}}$. 
We use the parameters $n=12$, 
$r_{\textrm{\scriptsize{ga}}}=0.5r_{\textrm{\scriptsize{cc}}}$, and 
$r_{\textrm{\scriptsize{cc}}}=3\sigma$ here. 
When the $i$th particle
has two or less particles within the cutoff distance 
$r_{ij}<r_{\textrm{\scriptsize{cc}}}$,
the particles could be on a plane, 
thereby making $\alpha_{\textrm{\scriptsize{pl}}} = 0$.

\subsection{Meshless spin membrane model}

In the meshless spin membrane model,
a tilt potential $U_{\rm {tilt}}$ and a bending potential $U_{\rm {bend}}$ are employed
instead of the aplanarity potential $U_\alpha$:  
 $U/k_{\rm B}T = \varepsilon (U_{\rm rep} + U_{\rm att} ) + U_{\rm {tilt}}+U_{\rm {bend}}$.
In this model, the spontaneous curvature of the membrane can be varied.
Each membrane particle has one orientation unit 
vector $\bm{u}_i\ (|\bm{u}_i| = 1)$. 
The potentials $U_{\textrm{\scriptsize{tilt}}}$ and $U_{\textrm{\scriptsize{bend}}}$ are given by
\begin{eqnarray}
U_{\textrm{\scriptsize{tilt}}}&=&  \frac{k_{\textrm{\scriptsize{tilt}}}}{2} 
\sum_{i<j} \left[ (\bm{u}_i\cdot\hat{\bm{r}}_{ij} )^2 + 
(\bm{u}_j\cdot\hat{\bm{r}}_{ij} )^2 \right] w_\textrm{\scriptsize{cv}} (r_{ij}), 
\nonumber \\
U_{\textrm{\scriptsize{bend}}}&=& \frac{k_{\textrm{\scriptsize{bend}}}}{2} \sum_{i<j} (\bm{u}_i -\bm{u}_j
- C_{\textrm{\scriptsize{bd}}} \hat{\bm{r}}_{ij} )^2 w_\textrm{\scriptsize{cv}}(r_{ij}),
\end{eqnarray}
where $\hat{\bm{r}}_{ij} = \bm{r}_{ij}/r_{ij}$.
These potentials are the discretized versions of
the tilt and the bending potentials of the tilt model \cite{hamm98,hamm00}, respectively.
The spontaneous curvature of the membrane 
can be controlled by the potential parameter $C_{\rm bd}$
as discussed later in Sec. \ref{sec:spcv}.
Recently, we employed similar potentials for a molecular lipid model
to form bilayer membranes \cite{nogu11}.
In the molecular model, the positions ${\bf r}^{\rm e}_i={\bf r}_i+{\bf u}_i \sigma$ were
used for calculation of the weight $w_\textrm{\scriptsize{cv}}(r_{ij}^{\rm e})$ 
to stabilize the bilayer structure.
In contrast, here,
the distance $r_{ij}$ between the centers of mass of the particles
is used to calculate $w_\textrm{\scriptsize{cv}}(r_{ij})$.
We use the same parameters in the functions 
$U_{\textrm{\scriptsize{rep}}}$, $U_{\textrm{\scriptsize{att}}}$,
and $w_\textrm{\scriptsize{cv}}(r_{ij})$ for both the models.
Unless otherwise specified,
we use $\varepsilon=4$.

\subsection{Simulation methods: Brownian dynamics}

We simulate the membranes in the NVT ensemble
(the particle number, volume, and 
temperature are kept constant).
The dynamics of the membrane is calculated using
Brownian dynamics (underdamped Langevin equation).
The motion of the membrane particles is given by
\begin{eqnarray} \label{eq:bd_cent}
m \frac{d^2\bm{r}_i}{dt^2} &=& -\zeta_{\rm G} \frac{d\bm{r}_i}{dt} +\bm{g}^{\rm G}_i(t) - 
\frac{\partial U}{\partial\bm{r}_i}, \\ \label{eq:bd_ori}
I \frac{d {\boldsymbol \omega}_{i}}{dt} &=&
 - \zeta_{\rm r} {\boldsymbol \omega}_i + \Big({\bf g}^{\rm r}_{i}(t)
 -\frac{\partial U}{\partial {\bf u}_{i}}\Big)^{\perp} + \lambda {\bf u}_{i},
\end{eqnarray} 
 where $m$ and $I$ are the mass and the moment of inertia of the molecule, respectively.
The angular velocity ${\boldsymbol \omega}_{i}= d{\bf u}_{i}/dt$
is rotated by the perpendicular force ${\bf f}^{\perp} ={\bf f}- ({\bf f}\cdot{\bf u}_{i}) {\bf u}_{i}$.
The length ${\bf u}_{i}^2=1$ is kept constant by a Lagrange multiplier $\lambda$.
The friction coefficients $\zeta_{\rm G}$ and $\zeta_{\rm r}$ and 
the Gaussian white noises ${\bf g}^{\rm G}_{i}(t)$ and ${\bf g}^{\rm r}_{i}(t)$
 obey the fluctuation-dissipation theorem;
\begin{eqnarray}
&& \langle g^{\beta_1}_{i,\alpha_1} (t) \rangle = 0, \\ \nonumber 
&& \langle g^{\beta_1}_{i,\alpha_1}(t) g^{\beta_2}_{j,\alpha_2} (t')\rangle = 2k_{\rm B}T\zeta_{\beta_1} \delta_{ij}\delta_{\alpha_1\alpha_2}\delta_{\beta_1\beta_2}
\delta (t-t').
\end{eqnarray}
Here, $\alpha_1, \alpha_2 \in \{ x,y,z\}$  and  $\beta_1, \beta_2 \in \{{\rm {G, r}}\}$.
In the following sections, we use the time unit
$\tau = \zeta_{\rm G}\sigma^2 /k_{\rm B}T$ and the energy 
unit $k_{\rm B}T$.
We use $m=\zeta_{\rm G} \tau$ and  $I=\zeta_{\rm r}\tau$.
For the mls or the spin model,
Eq. (\ref{eq:bd_cent}) or Eqs. (\ref{eq:bd_cent}) and  (\ref{eq:bd_ori})
 are integrated by the leapfrog 
algorithm with a time step of $\Delta t= 0.005\tau$, respectively.  
The simulations are performed with 
periodic boundary conditions in a box of dimensions
$L_x\times L_y \times L_z$.

\begin{figure}
\includegraphics{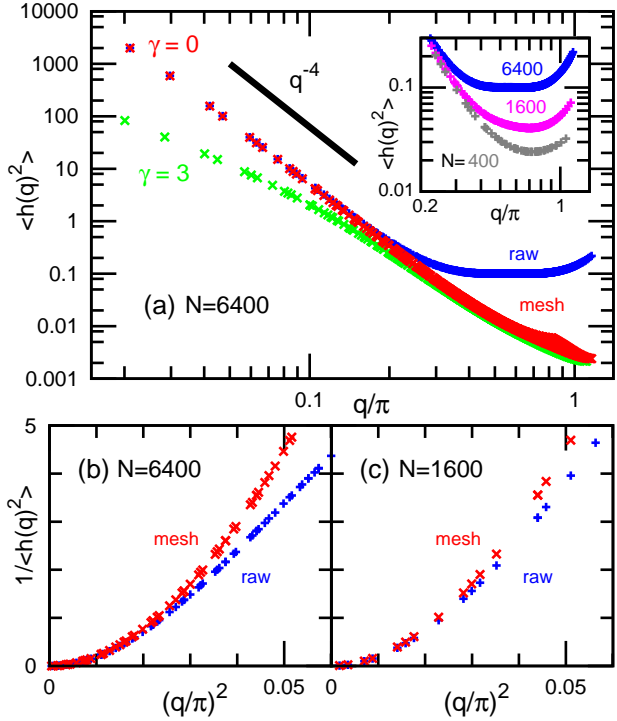}
\caption{\label{fig:hq_raw}
(Color online)
Spectra of undulation modes $\langle |h(q)|^2 \rangle$ of a planar membranes 
for the mls membrane model with $k_{\alpha}=10$ and $\varepsilon=4$.
Results for $\langle |h(q)|^2 \rangle$ calculated from the particle positions
(raw: $+$) and from the averaged positions on a square mesh (mesh: $\times$) are shown.
(a) Spectra for $\gamma=0$ ($A_{xy}/N\sigma^2=1.416$) and $\gamma=3$ ($A_{xy}/N\sigma^2=1.55$) at 
the number of particles $N=6400$.
The thick black line represents a slope of $q^{-4}$.
The inset shows $\langle |h(q)|^2 \rangle$ of raw data for a large $q$ at $N=400$, $1600$, and $6400$.
The dependence of $1/\langle |h(q)|^2 \rangle$ on $q^2$ for $N=6400$ and $N=1600$
is shown in (b) and (c), respectively. 
}
\end{figure}

\section{Thermal undulations of planar membranes}\label{sec:flat}

The undulation spectrum analysis of a planar membrane
is the most widely used method to estimate the bending
rigidity $\kappa$ in simulations. 
In this section, we compare fitting methods and establish a large 
dependence of the estimate value
on the cutoff frequency $q_{\rm {cut}}$.
The spectrum of the undulation
modes $\langle |h(q)|^2\rangle$ of the planar membranes in a Fourier
space is given by \cite{SafranBook,Goetz1999,lind00}
\begin{equation}
\langle |h(q)|^2\rangle=\frac{k_{\rm B}T}{\gamma q^2 + \kappa q^4}.
\label{eq:hq0}
\end{equation}
We calculate $|h(q)|^2$ for the planar membranes with $L_x=L_y$
 from the raw data (the particle position ${\bf r}_i$)
as well as from the positions  averaged on 
a $\sqrt{N/2} \times \sqrt{N/2}$ square mesh with 
$(x_{\rm {mh}}, y_{\rm {mh}})=(d_{\rm {mh}} n_x, d_{\rm {mh}} n_y)$.
The height $z_{\rm {mh}}$ of a mesh point is obtained from the weighted 
average of the molecular position ${\bf r}_i$ in the four neighbor cells with
$z_{\rm {mh}}=  \sum_i z_i w_{\rm {mh}}(x_i, y_i)/(\sum_i w_{\rm {mh}}(x_i, y_i))$
and $w_{\rm {mh}}(x_i, y_i) = 
(1-|x_i-x_{\rm {mh}}|/d_{\rm {mh}})(1-|y_i-y_{\rm {mh}}|/d_{\rm {mh}})$.
We refer to the former and the latter spectra 
as the  ``raw spectrum'' and the ``mesh spectrum'' respectively.
Figure~\ref{fig:hq_raw}(a) clearly shows the $q^{-4}$ dependence of the 
tensionless membrane (surface tension $\gamma=0$).
For increasing $\gamma$, $\langle |h(q)|^2\rangle$ decreases at
a low frequency $q$ 
(compare the data at $\gamma=0$ and $\gamma=3$ in Fig.~\ref{fig:hq_raw}(a)).
The mesh spectrum approaches zero at a high $q$.
On the contrary, the raw spectrum saturates at a finite value, 
which increases with increasing $N$
(see the inset of Fig.~\ref{fig:hq_raw}(a)).
These high $q$ modes are caused by particle protrusion
due to the short-range potential interactions between particles.
Averaging over the mesh removes most of the effects of these particle
protrusions at a high $q$.
The effects of protrusions on bending rigidity 
estimation will be discussed in the last part of this section,
with mls and spin models.

\begin{figure}
\includegraphics{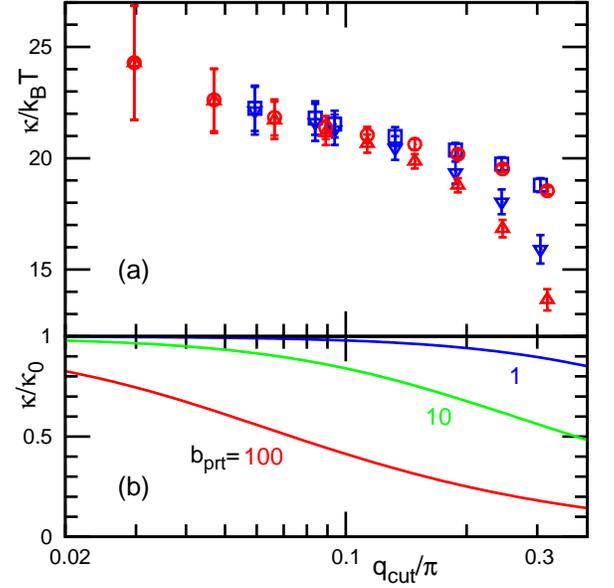}
\caption{\label{fig:hq_log}
(Color online)
Estimation of the bending rigidity $\kappa$ of the tensionless membrane 
from the log-log fit for (a) the mls membrane model with $k_{\alpha}=10$ 
and (b) the phenomenological function Eq. (\ref{eq:hq_pheno}).
(a) The symbols 
 (blue or dark gray: $\Box, \triangledown$) and 
(red or light gray: $\circ, \triangle$)
represent the data for $N=1600$ and $6400$, respectively.
The symbols at relatively lower positions 
($\triangledown, \triangle$) and the symbols 
at relatively higher positions ($\Box, \circ$) 
represent the data estimated from the fits for the raw  spectrum
and the mesh spectrum, respectively.
}
\end{figure}

For the estimation of $\kappa$, two types of fits are widely used.
(i) a log-log fit for the tensionless membranes,
\begin{equation}
\ln(\langle |h(q)|^2\rangle)= -\ln\Big(\frac{\kappa}{k_{\rm B}T}\Big) - 4\ln(q),
\label{eq:hq_log}
\end{equation}
and (ii) an inverse power-spectrum fit,
\begin{equation}
\frac{1}{\langle |h(q)|^2\rangle} = \frac{\gamma q^2 + \kappa q^4}{k_{\rm B}T}.
\label{eq:hq_inv}
\end{equation}
In the latter fit, $\gamma$ can also be estimated.
We calculated both of them using the linear least squares fit
for various cutoff values given by $q_{\rm {cut}}$.
For the fits with one fit parameter ($\kappa$),
 the surface tension calculated from the pressure tensor
is used as the value of  $\gamma$.
The surface tension is given by
\begin{equation}
\gamma= \langle P_{zz} - (P_{xx} + P_{yy})/2\rangle L_z,
\label{eq:stpt}
\end{equation} 
with the diagonal components of the pressure tensor
\begin{equation}
\label{eq:pressure_tensor}
P_{\alpha\alpha} = (Nk_{\rm B}T - 
     \sum_{i} \alpha_{i}\frac{\partial U}{\partial {\alpha}_{i}} )/V,
\end{equation} 
where $\alpha \in \{x, y, z\}$ \cite{rowl82,alle87}.
In calculating $P_{\alpha\alpha}$, 
 the periodic image $\alpha_i + nL_\alpha$
nearest to the other interacting particles is employed,
when the potential interaction crosses the periodic boundary.
We compare these two surface tensions,
calculated from the pressure and the thermal undulations,
later in this section.

Figure \ref{fig:hq_log}(a) shows the bending rigidity $\kappa$ estimated from the log-log fit of
Eq. (\ref{eq:hq_log}) for the data for which $q<q_{\rm {cut}}$.
As the cutoff frequency $q_{\rm {cut}}$ increases,
the estimate value of $\kappa$ gradually decreases.
The fits to the mesh spectrum are less sensitive to $q_{\rm {cut}}$ than those to the raw spectrum.
This $q_{\rm {cut}}$ dependence is caused by the neglected fluctuation modes.
As shown in Fig.~\ref{fig:hq_raw}(a), 
$\langle |h(q)|^2\rangle$ of the raw spectrum deviates from $q^{-4}$ at a high $q$
because of particle protrusions.
Goetz {\it et al.} reported that the protrusion modes of lipid molecules 
have a $q^{-2}$ dependence \cite{Goetz1999}.
To clarify the influence of these protrusion modes on the spectrum, 
we test the least squares fit
to a phenomenological function
\begin{equation}
\langle |h(q)|^2\rangle = \frac{k_{\rm B}T}{\kappa_0}\Big(\frac{1}{(\gamma_0/\kappa_0) q^2 + q^4} 
+ \frac{b_{\rm {prt}}}{q^2} \Big).
\label{eq:hq_pheno}
\end{equation}
The first term is 
 the thermal-undulation mode with
 the bending rigidity $\kappa_0$  and surface tension $\gamma_0=0$,
and the last term ($\propto q^{-2}$) is  a protrusion mode.
Figure \ref{fig:hq_log}(b) shows $\kappa$
estimated by the log-log fit of Eq. (\ref{eq:hq_inv}) to the 
phenomenological function of Eq. (\ref{eq:hq_pheno}) with $\gamma_0=0$
 for $0.01\pi<q<q_{\rm {cut}}$.
At a high $b_{\rm {prt}}$ or high $q_{\rm {cut}}$,
the fit gives a lower $\kappa$  than the actual bending rigidity $\kappa_0$.
It qualitatively reproduces the $q_{\rm {cut}}$ dependence obtained in the simulations.
A similar decrease is observed when $b_{\rm {prt}}/q^{-1}$ is used instead of $b_{\rm {prt}}/q^{-2}$.
Thus, this dependence is not sensitive to the function shape.
As a result, the fit with a lower $q_{\rm {cut}}$ should yield a more accurate $\kappa$
close to $\kappa_0$. However, 
since the number of the data points for the fits is fewer for lower $q_{\rm {cut}}$
and the spectrum at low $q$ has a larger statistical error,
the error bar is larger in low $q_{\rm {cut}}$ region. 
Thus, the medium cutoff ($q_{\rm {cut}} \simeq 0.1\pi$ in Fig. \ref{fig:hq_log}) 
is a reasonable choice for the estimation of $\kappa$ from the log-log fit.

\begin{figure}
\includegraphics{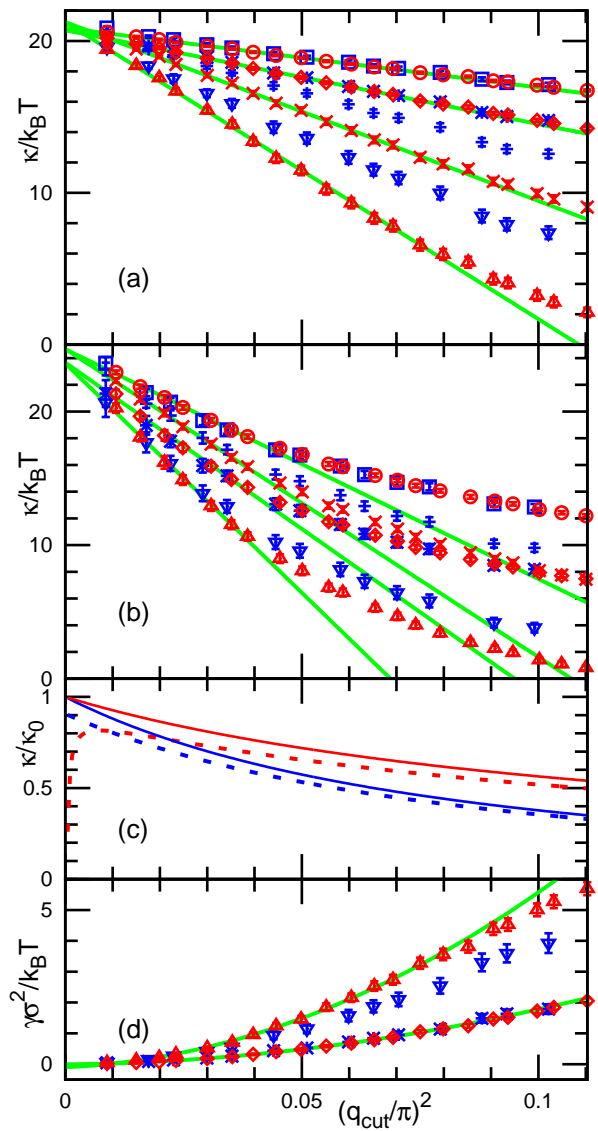}
\caption{\label{fig:hq_cut}
(Color online)
Estimation of the bending rigidity $\kappa$ 
from the fit to $1/\langle |h(q)|^2 \rangle$ with Eq. (\ref{eq:hq_inv})
for (a) the mls model with $k_{\alpha}=10$, (b) the spin model with 
$k_{\rm {bend}}=k_{\rm {tilt}}=15$ and $C_{\rm {bd}}=0$ at $\varepsilon=4$,
and (c) the phenomenological function in Eq. (\ref{eq:hq_pheno}).
The symbols (blue or dark gray: $\Box, \ast, +, \triangledown$ and 
red or light gray: $\circ, \diamond, \times, \triangle$)
represent the data for $N=1600$ and $6400$, respectively.
From the top to the bottom,
the one-parameter fit for a mesh spectrum ($\Box, \circ$);
two-parameter fit for a mesh spectrum ($\ast, \diamond$);  
one-parameter fit for a raw spectrum ($+, \times$); and
two-parameter fit for a raw spectrum ($\triangledown, \triangle$)
are shown. 
The solid lines in (a) and (b) are obtained 
by a least squares fit for the data in (a) $(q_{\rm {cut}}/\pi)^2<0.08$ and (b) $(q_{\rm {cut}}/\pi)^2<0.04$.
(c) The solid and dashed lines represent the data
for $\gamma_0/\kappa_0=0$ and $0.05$ at $b_{\rm {prt}}=1$, respectively.
The upper or lower lines show the results obtained using
 one- ($\kappa$) or two- ($\kappa$, $\gamma$) parameter fits, respectively.
(d) The surface tension estimated from the two-parameter fit for the mls model.
The solid lines represent the curves fitted by 
$\gamma=\gamma_{\rm {fl}} + a_{\gamma} q_{\rm {cut}}+ b_{\gamma} q_{\rm {cut}}^2$
for the data at $(q_{\rm {cut}}/\pi)^2 <0.08$. 
}
\end{figure}

\begin{figure}
\includegraphics{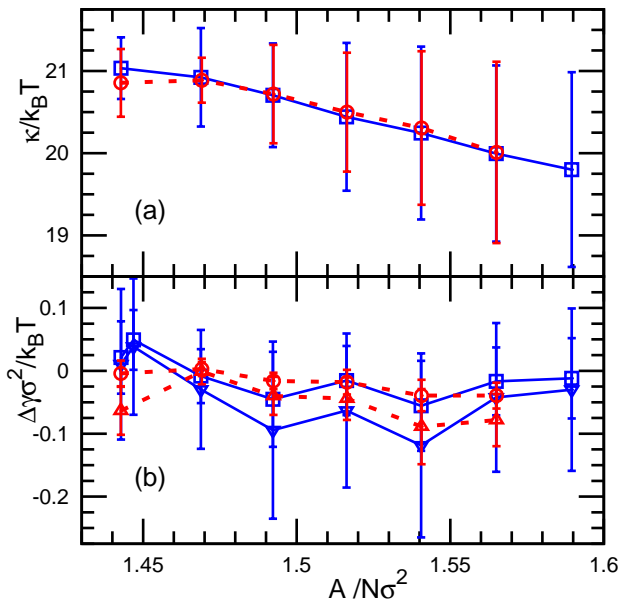}
\caption{\label{fig:kappa_ar}
(Color online)
Intrinsic area $A$ dependence of (a) the bending rigidity $\kappa$ and 
(b) the difference $\Delta\gamma=\gamma_{\rm {fl}}-\gamma_{\rm {pr}}$ of 
the surface tensions estimated from the undulations
(Eq.~(\ref{eq:hq_inv})) and the pressure tensor (Eq.~(\ref{eq:stpt})).
The solid line with ($\Box, \triangledown$) and dashed line with ($\circ, \triangle$)
represent the data for $N=1600$ and $6400$, respectively.
(b) The symbols ($\triangledown, \triangle$) and ($\Box, \circ$) 
represent the data estimated from the raw spectrum and the
mesh spectrum, respectively.
}
\end{figure}

Next, we explain the inverse power-spectrum fit of Eq. (\ref{eq:hq_inv}).
This fit also shows a large dependence on $q_{\rm {cut}}$.
Four types of fits (with one or two fit parameters for the raw and mesh spectra) 
for the tensionless membranes are shown in Fig. \ref{fig:hq_cut}.
The spectra are fitted with one ($\kappa$) or two ($\kappa, \gamma$) fit parameters.
The one-parameter fit provides a lower slope of the $q_{\rm {cut}}$--$\kappa$ curve
than the two-parameter fit.
The mesh spectrum gives lower slopes for $\kappa$ and $\gamma$ 
than the raw  spectrum does (see Figs. \ref{fig:hq_cut}(a) and (d)).
Thus, substantial differences are seen between 
the $\kappa$ values estimated by different fits
for a finite $q_{\rm {cut}}$.
However, all the fits converge at $q_{\rm {cut}} \to 0$.
A similar $q_{\rm {cut}}$  dependence is observed for the fit to the phenomenological function
in Eq. (\ref{eq:hq_pheno}) with $\gamma_0=0$.
At $q_{\rm {cut}} \to 0$, $\kappa$ values converge to the correct value $\kappa_0$.
Therefore, the bending rigidity $\kappa$ can be
estimated by an extrapolation using the linear least squares fit to a straight line
(see solid lines in Figs. \ref{fig:hq_cut}(a) and (b)).

We select the extrapolated value at $q_{\rm {cut}}=0$
from a one-parameter fit for the mesh spectrum as the bending rigidity $\kappa$,
considering that it has the lowest slope of the $q_{\rm {cut}}$--$\kappa$ curve.
We estimate the numerical error in $\kappa$ from two contributions; {\it i.e.}, 
for $\Delta \kappa=\Delta \kappa_{\rm {f0}} + \Delta \kappa_{\rm {f1}}$:
the maximum difference $\Delta \kappa_{\rm {f0}}$ between
four extrapolated values is considered as the numerical error size from the 
choice of functions, and $\Delta \kappa_{\rm {f1}}$ is the error of the least squares fit.

As the membrane area increases, the surface tension increases.
We investigated the area dependence of $\kappa$ using the  extrapolation method
for the parameters ($k_{\alpha}=10$, $\varepsilon=4$) 
investigated in our previous paper \cite{Nog06PRE}.
The intrinsic area $A$ of the membrane is larger than the projected area 
$A_{xy}$ in the $xy$ plane because of the membrane undulations.
We calculate $A$ from the mesh points for the mesh spectrum.
Figure \ref{fig:kappa_ar} shows the area dependence of $\kappa$  and 
the difference $\Delta\gamma=\gamma_{\rm {fl}}-\gamma_{\rm {pr}}$ in
the surface tension estimated from the two methods.
The surface tensions $\gamma_{\rm {pr}}$ are estimated from the pressure tensor (Eq.~(\ref{eq:stpt}));
$\gamma_{\rm {pr}}=0$ and $A/N\sigma^2=1.443$ at $A_{xy}/N\sigma^2=1.416$, 
whereas $\gamma_{\rm {pr}}=2.96$ and $A/N\sigma^2=1.565$ at $A_{xy}/N\sigma^2=1.55$.
The surface tension estimated by the  extrapolation with 
$\gamma =\gamma_{\rm {fl}} +a_{\gamma} q_{\rm {cut}}+b_{\gamma} q_{\rm {cut}}^2$ 
has a very good agreement with $\gamma_{\rm {pr}}$,
as shown in Fig. \ref{fig:kappa_ar}(b).
Previously, Imparato reported that the surface tension
estimated from the thermal undulations 
is smaller than that estimated from the pressure tensor in 
molecular simulations \cite{impa06}. The reported difference may be 
due to the similar effects of a finite $q_{\rm {cut}}$.

Although the estimated $\kappa$ decreases with an increase in $A$,
it is accompanied by larger error bars.
Thus, we do not find a clear dependence of $\kappa$ on $\gamma$.
For a finite $\gamma_0$,
the fit for the  phenomenological function of Eq. (\ref{eq:hq_pheno})
shows a deviation from the correct value $\kappa_0$ at $q_{\rm {cut}} \to 0$
(see  Fig. \ref{fig:hq_cut}(c)).
The one-parameter fit shows an abrupt decrease at $q_{\rm {cut}}\simeq 0$,
whereas rapid decreases are not observed in the simulations.
These results suggest that systematic errors 
may be involved in the estimation of $\kappa$ at a finite $\gamma$.
Further investigations are needed to clarify the $\kappa$ dependence on $\gamma$.

The extrapolation method works well for  the spin membrane model as well as the mls membrane model.
The spin model has a larger $q_{\rm {cut}}$ dependence.
The $q_{\rm {cut}}$--$\kappa$ curve deviates
from the straight line at 
$(q_{\rm {cut}}/\pi)^2 \gtrsim 0.05$ and $(q_{\rm {cut}}/\pi)^2 \gtrsim 0.1$ 
for the spin and mls models, respectively
(compare Figs. \ref{fig:hq_cut}(a) and (b)).
This suggests that the spin model has greater particle protrusion than the mls model.
Because the protrusion is induced by the short range interactions between particles or molecules,
it is sensitive to the potentials of simulation models.
To use this method, one should ensure
that the spectrum at a sufficiently low $q$ is included for the linear extrapolation.
The bilayer membranes of the spin molecular model \cite{nogu11}
show a similar dependence as that of the meshless spin model (data not shown).
Simulation models accompanied with larger protrusions require larger system sizes 
to estimate $\kappa$ from the thermal undulations.

For $\kappa$ extrapolated at $q_{\rm {cut}} \to 0$,
no significant dependence on the system size $N$ is detected.
All of the extrapolations for $N=1600$ and $6400$ converge
(see Figs. \ref{fig:hq_cut} and \ref{fig:kappa_ar}).
At a finite $q$, the raw spectrum is dependent on $N$
because the protrusion amplitude increases with increasing $N$
(compare Figs. \ref{fig:hq_raw}(b) and (c)).
The slope of the  $q_{\rm {cut}}$--$\kappa$ curve is higher at a larger $N$
for the raw spectrum (see Fig. \ref{fig:hq_cut}).

We also estimated $\kappa$ using a nonlinear least squares fit for 
$\langle |h(q)|^2\rangle$ without and with the protrusion term $\sim q^{-2}$
[Eq. (\ref{eq:hq0}) and Eq. (\ref{eq:hq_pheno})].
Although these fits are not sensitive to $q_{\rm {cut}}$, 
they have larger errors than the above fitting methods. 
Therefore, we conclude that the inverse power
spectrum with an extrapolation to $q_{\rm {cut}} \to 0$
is the best fitting method for the undulation of planar membranes.
In the Appendix, we list the bending rigidity $\kappa$ 
estimated by the inverse power-spectrum fit
for the mls and spin membrane models with various parameters.

\section{Tubular membranes with no spontaneous curvature}\label{sec:nocv}

\begin{figure}
\includegraphics[width=0.9\linewidth]{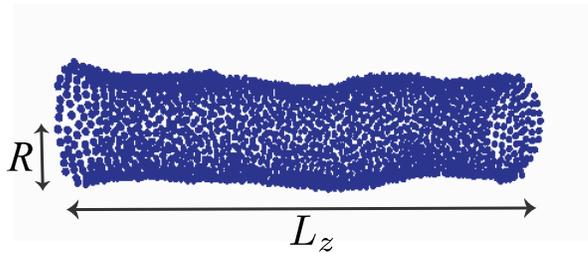}
\caption{\label{fig:tube} (Color online)
Snapshot of a tubular membrane in the simulation of
the mls membrane model at $N=2400$, $k_\alpha = 10$, $\varepsilon=4$, and $L_z=80\sigma$.
}
\end{figure}

In this section, we present 
the estimation of the bending rigidity $\kappa$ from tubular membranes.
For a tubular membrane with a 
radius $R$ and a length $L_z$,
the curvature free energy Eq. (\ref{eq:Helfrich}) is 
written as
\begin{equation} \label{eq:F_tube}
\mathcal{F} = 2\pi RL_z \left[ \frac{\kappa}{2}\left( \frac{1}{R}
-C_0 \right)^2 \right].
\end{equation}
Under the fixed area condition $A = 2\pi RL_z = \textrm{const.}$
the axial force  $f_z=\partial F/\partial L_z|_A$ is given by
\begin{equation}
f_z = 2\pi\kappa \Big(\frac{1}{R} - C_0\Big). \label{eq:tension}
\end{equation}
lateral tension is anisotropic:
 $\gamma_z=f_z/2\pi R=\kappa (1/R-C_0)$ in the axial direction,
$\gamma_{\theta}=0$  in the azimuth direction, 
and $\gamma_{\rm {av}}= \gamma_z/2$ in average.
Although we assume the constant area here,
the area compressibility does not change the force,  Eq. (\ref{eq:tension}).
When the area compressibility is taken into account,
the free energy Eq. (\ref{eq:F_tube}) has an additional term $U_{\rm {ar}}(A)$
($U_{\rm {ar}}(A) = K_{\rm A}(A-A_0)^2/2A_0$ for $A-A_0 \ll 1$,
where $A_0$ is the area of the tensionless membrane).
Then, the same force is derived
from $f_z=\partial F/\partial L_z$ and $\partial F/\partial R=0$.
At $C_0=0$,
the force is inversely proportional to $R$ ($f_z = 2\pi\kappa/R$). 
Using this relation, $\kappa$ was previously estimated in experiments \cite{Bo,Evans1990,Dai,Evans1996,Cuvelier}
and molecular simulations \cite{Harmandaris}.

In this section, we investigate stretched cylindrical membranes 
with $C_0=0$
using the mls membrane model
(see Fig. \ref{fig:tube}).
 All the tubes are connected periodically 
in the axial direction with the periodic length $L_z$.
The initial conformations at each $L_z$
are made by slow stretching or shrinkage
of the length $L_z$ with a speed 
less than $dL_z /dt = 0.002\sigma /\tau$. 
We checked that no hystereses are seen in results between stretching and shrinkage.
The bending rigidity $\kappa$ is measured
at fixed $L_z$. After discarding the data for
the first calculation period $1600\tau$, the data are averaged
for a period $48000\tau$ ($72000\tau$) for $N=2400$ ($1200$). 
Eight simulations  starting with
 independent initial conformations are performed.

\begin{figure}
\includegraphics{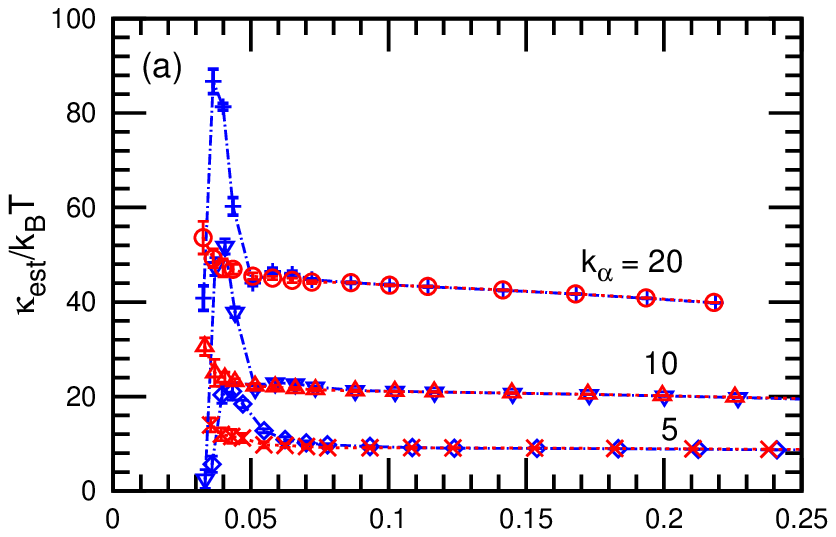}
\includegraphics{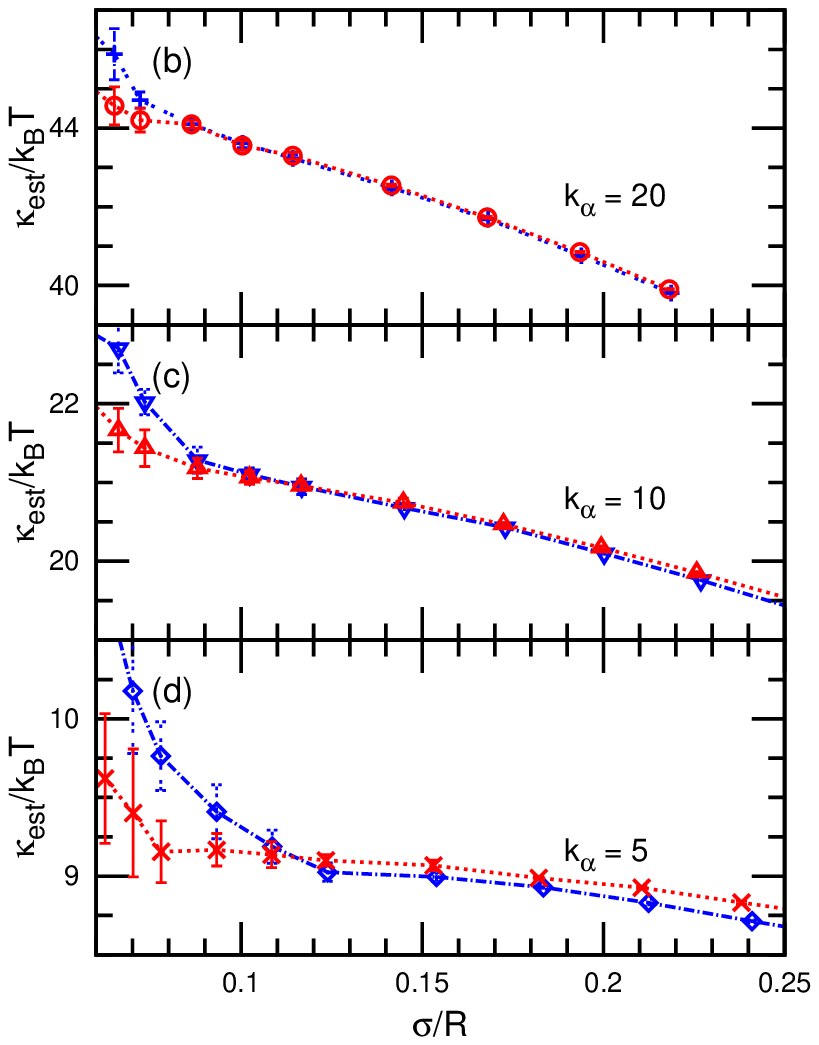}
\caption{\label{fig:kappa_tube} (Color online)
(a) Bending rigidity 
${\kappa}_{\textrm{\scriptsize{est}}} = f_z R/2\pi$ 
estimated from the stretching force measurement 
using the mls membrane model. 
The symbols (blue or dark gray: $+,\triangledown ,\diamond$ 
and red or light gray: $\circ,\triangle, \times$)
represent the data at $N=1200$ and $2400$, 
respectively, for $k_\alpha =20$ ($+, \circ$);
$10$ ($\triangledown, \triangle$);
and $5$ ($\diamond,\times$). 
(b), (c), and (d) show magnified plots of the data
shown
in (a) for $k_\alpha = 20$, $10$, and $5$, respectively. 
}
\end{figure}

Figure \ref{fig:kappa_tube} shows the estimate values of
the bending rigidity ${\kappa}_{\textrm{\scriptsize{est}}} = f_z R/2\pi$ 
for various $L_z$.
The inverse radius of the tube $\sigma /R$
is employed for the horizontal axis.
With increasing $\sigma /R$,
the cylinder tube becomes narrower and longer.
The radius $R$ of the cylinder is simply estimated
by averaging the distance of each particle 
from the cylindrical axis:
$R= \langle  \sum_i\{(x_i-x_{\rm G})^2+(y_i-y_{\rm G})^2\}^{1/2}/N \rangle$,
where $(x_{\rm G}, y_{\rm G})$ is the center of mass of the membrane
projected on the $xy$ plane.
The estimate values of $\kappa_{\textrm{\scriptsize{est}}}$
for  long tubes ($\sigma /R \gtrsim 0.06$)
have very good agreements with those obtained for
planar membranes in Sec. \ref{sec:flat}.

For shorter tubes ($\sigma /R \lesssim 0.06$),
large bumps (or peaks) appear at $N=1200$ (see Fig. \ref{fig:kappa_tube}(a)).
Because they are suppressed at $N=2400$ 
(tubes that have twice the length for the same radius),
these bumps are likely caused by the finite size effects in the $z$ direction. 
For long tubes,
$\kappa_{\rm est}$ decreases slightly with increasing $\sigma/R$.
This may show the dependence of $\kappa$ on the area expansion
or be on account of the higher-order terms of the bending elasticity 
discussed in Ref. \cite{Harmandaris}.
As  explained later in this section,
the intrinsic area $A$ is larger for longer tubes.
The ${\kappa}_{\textrm{\scriptsize{est}}}$ decrease resembles
that seen in the estimation for
the planar membranes of Fig. \ref{fig:kappa_ar}.
The decrease rate strongly depends on $k_\alpha$;
$- d({\kappa}_{\textrm{\scriptsize{est}}}/ k_{\rm B}T)/d (\sigma/R) =3$, $11$, and $31$ 
for $k_\alpha = 5$, $10$, and $20$, respectively.
This difference can be partially explained by the effects of the thermal fluctuations.
Recently, Barbetta {\it et al.} \cite{Fournier2009} 
derived the axial force under the thermal fluctuations
using perturbation theory.
In their theory, the force is given by 
\begin{equation}
\frac{f_z R}{2\pi}= \kappa - \frac{k_{\rm B}T}{2\pi^2} R^2\Lambda^2,
\end{equation}
where $\Lambda$ is the cutoff wave vector.
For $\Lambda=1/2\sigma$,
 ${\kappa}_{\textrm{\scriptsize{est}}}=f_z R/2\pi$ increases by $k_{\rm B}T$
from $\sigma/R=0.1$ to $0.25$.
Thus, the decrease rate for $k_{\alpha}=5$ in Fig. \ref{fig:kappa_tube}
is reduced by the thermal-fluctuation effects. 

\begin{figure}
\includegraphics{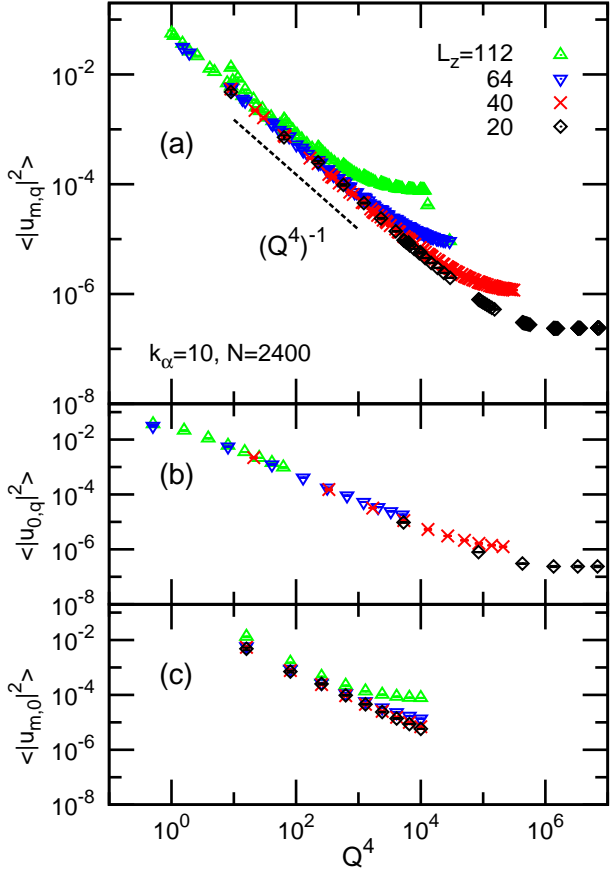}
\caption{\label{fig:fluct_umn} (Color online)
(a) Undulation spectra
$\langle |u_{m,\overline{q}} |^2 \rangle$
of the tubular membranes of the mls model at 
$N=2400$ and $k_\alpha = 10$, for 
$L_z/\sigma = 112\ (\triangle),\ 64\ 
(\triangledown),\ 40\ (\times)$, and 
$20\ (\diamond)$. 
The data for $0\le m,n\le 10$ are 
shown.
The horizontal axis is the 
fourth power of the effective frequency $Q$ defined in 
Eq. (\ref{eq:wvn_cyl}).
(b)(c) Spectra  for (b) $m=0$ and $\ 0\le n\le 10$ 
and (c) $0\le m\le 10$ and $n=0$
are extracted from the data in (a) to clearly show
the $m$ and $\overline{q}$ dependence. 
}
\end{figure}

The bending rigidity $\kappa$
can also be estimated from the surface 
undulations of tubular membranes.
Recently, Fournier {\it et al.}  
studied the thermal undulations
on a cylindrical membrane theoretically \cite{Fournier2007}.
They predicted the nontrivial effects of the critical Goldstone modes
for narrow and long tubes.
We numerically analyze the surface fluctuations
and membrane area in comparison with their 
theoretical framework.
The membrane position is expressed in the cylindrical coordinates
$\bm{r} (\theta, \zeta )/R = (
 [ 1+u(\theta, \zeta) ] \cos \theta,
[ 1+u(\theta, \zeta) ] \sin \theta,
\zeta)$
for $0\le \theta <2\pi$ and $0 \le \zeta < L_z/R$.
We calculate $u(\theta, \zeta)$
from the 
raw data of the particle positions $\bm{r}_i$. 
The cylindrical axis $(x_{\rm G}, y_{\rm G})$ 
and the radius $R$ are 
estimated in the same manner as in the 
above force measurement.
The Fourier 
modes of the cylindrical surface fluctuations
are given by
\begin{equation}
u (\theta, \zeta) = \sqrt{\frac{R}{2\pi L_z}}
\sum_{m,\overline{q}} u_{m,\overline{q}}
e^{i (m\theta + \overline{q}\zeta )},
\end{equation}
where $\overline{q}=2\pi nR/L_z$,
 $|m| \le 2\pi R/l $,
and $|n| \le L_z / l$.
The cutoff length $l$ is
the mean distance between neighboring membrane particles 
in meshless membrane models
or the membrane thickness in molecular models.

At thermal equilibrium, 
surface undulation of the cylindrical membrane
can be estimated by the perturbation theory.
The spectrum is given by \cite{Ouyang1989,Fournier2007}
\begin{eqnarray}
\langle |u_{m,\overline{q}} |^2 \rangle
&=& \frac{k_{\rm B}T}{\kappa Q^4},
\label{eq:fluctuation} \\
Q^4 &=&  (m^2-1)^2 + \overline{q}^2
 (\overline{q}^2 +2m^2 ), \label{eq:wvn_cyl}
\end{eqnarray}
where $Q$ denote the normalized amplitude of frequencies 
in the two-dimensional cylindrical space.
Since the expression Eq. (\ref{eq:fluctuation}),
for cylindrical membranes,
is the counterpart of Eq. (\ref{eq:hq0}) 
for planar membranes,
$\kappa$ can be estimated using a similar fitting method.

Figure \ref{fig:fluct_umn} shows the undulation spectra
for $L_z/\sigma=20$, $40$, $64$, and $112$
($R/\sigma= 27.1, 13.6, 8.58$, and $5.01$)
at $N=1200$ and $k_\alpha = 10$. 
The modes for small frequencies at $0\le m,n\le 10$
are shown here ($\langle |u_{1,0}|^2 \rangle$ 
are omitted because of their divergence).
While the amplitudes
$\langle | u_{m,n}|^2\rangle$ 
at a low $\overline{q}$ exhibit deviations
for the narrow tubes ($L_z/\sigma=112$),
they clearly show $Q^{-4}$ dependence 
at a low $Q$ when
the ratio between
the circumferential length and the
cylinder length $2\pi R/L_z$ is around unity.
In these regions, $\kappa$ can be estimated by
least squares fits, as explained in Sec. \ref{sec:flat}.

\begin{figure}
\includegraphics{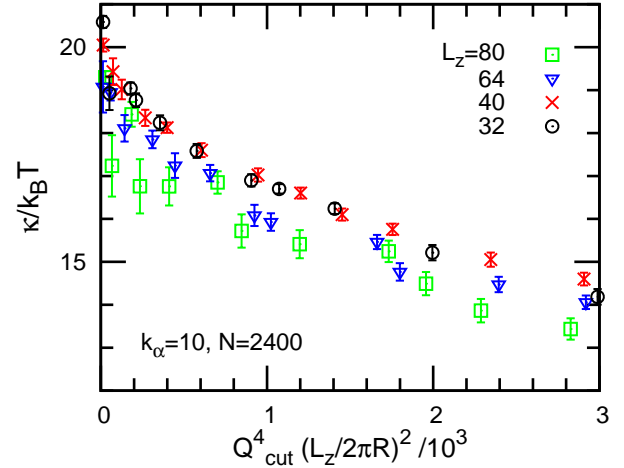}
\caption{\label{fig:qcutfit} (Color online)
Estimation of the bending rigidity $\kappa$
for the mls membrane model 
from the fit to $1/\langle |u_{m,\overline{q}}|^2 \rangle$
with Eq. (\ref{eq:fluctuation}), 
for $L_z/\sigma=80\ (\square),\ 64\ (\triangledown),
\ 40\ (\times),$ and 32 $(\circ)$, for various 
cutoff frequencies $Q_{\rm {cut}}^4$.
The data are fitted for
$0\le m,n\le 10$ and $Q^4\le Q_{\rm {cut}}^4$.
}
\end{figure}

Figure \ref{fig:qcutfit} shows
 the bending rigidity 
$\kappa$ obtained by a linear least squares fit
with various upper cutoffs $Q_{\rm {cut}}^4$
of the inverse power spectrum  
$\langle | u_{m,\overline{q}}|^2\rangle^{-1}$ 
using Eq. (\ref{eq:fluctuation}), 
in a manner similar to that used in Sec. III.
When the horizontal axis is normalized by $(2\pi R/L_z)^2$,
the data for all $L_z$ overlap.
This dependence is very similar
to that of the planar membranes shown in 
Fig. \ref{fig:hq_cut}(a). $\kappa$ 
approaches a value of around $20$ as $Q_{\rm {cut}}\to 0$. 
Thus, the spectra of $\langle |u_{m,\overline{q}}|^2\rangle$ 
well reflect the bending rigidity of the membranes.

\begin{figure}
\includegraphics{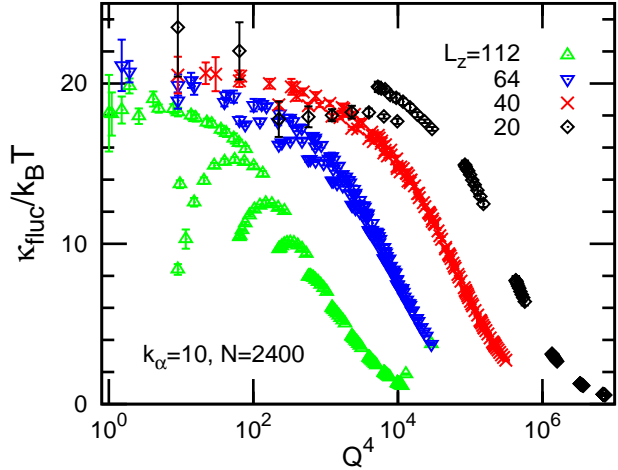}
\caption{ 
\label{fig:sp_cyl} (Color online)
Dependence of estimate values of the bending rigidity
$\kappa_{\textrm{\scriptsize{fluc}}}$ from the 
surface fluctuation spectrum 
on the frequency $Q^4$,
for $L_z/\sigma=112$ ($\triangle$), $64$ ($\triangledown$),
$40$ ($\times$), and $20$ ($\diamond$), 
at $k_\alpha =10$ and $N=2400$.  
 }
 \end{figure}

\begin{figure}
\includegraphics{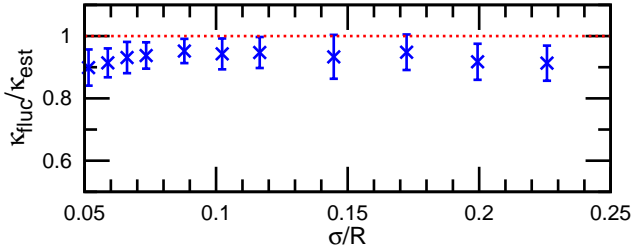}
\caption{\label{fig:comparison} (Color online)
Ratio $\kappa_{\textrm{\scriptsize{fluc}}} / 
\kappa_{\textrm{\scriptsize{est}}}$ between
the estimate values of the bending rigidity 
 from the force measurement Eq. (\ref{eq:tension}) 
and from the surface fluctuation spectrum shown
in Fig. \ref{fig:sp_cyl} at
$k_\alpha = 10$ and $N=2400$.
The lowest ten $Q^4$ modes
are used to estimate $\kappa_{\rm fluc}$.
}
\end{figure}

Figure \ref{fig:sp_cyl} shows the dependence of 
estimate values of 
the bending rigidity $\kappa_{\textrm{\scriptsize{fluc}}}
= k_{\rm B}T /\langle | u_{m,\overline{q}} |^2\rangle Q^4$
on the frequency ($Q^4$) using Eq. (\ref{eq:fluctuation}).
In this figure, we  plot the data for  $0\le m,n\le 10$. 
In intermediate length scales, where the 
radius of the tube is $0.045\lesssim \sigma /R\lesssim 0.1$,
the spectrum of $\kappa_{\textrm{\scriptsize{fluc}}}$ collapses
into a smooth shape, whereas 
systematic deviations for a specific $m$ or
$\overline{q}$ appear
at $L_z/\sigma=112$ and $20$.
Low $Q$ values of $\kappa_{\rm fluc}$, which 
represent longer-wavelength properties, 
 well converge to a value around $\kappa_{\rm est}$. 
To compare $\kappa_{\textrm{\scriptsize{fluc}}}$ 
with $\kappa_{\textrm{\scriptsize{est}}}$,
their ratio $\kappa_{\textrm{\scriptsize{fluc}}}/
\kappa_{\textrm{\scriptsize{est}}}$ is shown in Fig. \ref{fig:comparison}.
In the region 
plotted in Fig. \ref{fig:comparison} ($15\lesssim L_z/\sigma 
\lesssim 25$), 
the ratio is constant $\simeq 0.9$,
whereas both of them decrease with increasing $\sigma/R$.
These tendencies 
are also obtained for other $k_\alpha$ or $N$.
The results could be reflecting the dependence of 
$\kappa$ on the intrinsic area $A$ for 
the meshless model. 

\begin{figure}
\includegraphics{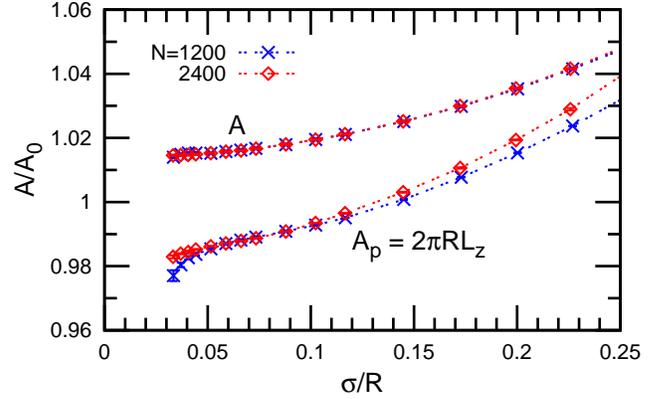}
\caption{\label{fig:cyl_area} (Color online)
Tube radius dependence of the intrinsic membrane area $A$ 
and the projected area $A_{\rm p} = 2\pi RL_z$ 
at $k_\alpha=10$. 
The area is
normalized by that of the tensionless 
planar membrane ($A_0/N\sigma^2 = 1.443$).
Blue (dark gray) and red (light gray) points show the 
results for $N=1200$ and $2400$, respectively. }
\end{figure}

In the meshless membranes, the 
tubular membrane area is slightly expanded owing to
the axial tension. 
Here, we estimate the membrane 
area in the following manner:
A Delaunay tessellation is performed
for the ($\theta, \zeta$) coordinates
to construct a triangulated 
surface on the membrane. 
Then, 
the intrinsic  membrane area $A$ is calculated
as the sum over the area of the triangles for the 3D particle positions. 
Figure \ref{fig:cyl_area} shows the intrinsic area $A$ for 
$N=1200$ and $2400$ in comparison with 
the projected area $A_{\rm p} = 2\pi RL_z$. 
As the membrane area is expanded for larger axial tension 
($\gamma_z = \kappa /2R^2$), 
$A$ and $A_{\rm p}$ approach each other.
The area expansion of $A$ is more than twice 
that of the planar membranes for the same average surface tension
$\gamma_{\rm {av}}$.
 The anisotropy of the surface tension 
results in a low effective area compression modulus $K_{\rm A}$.

\begin{figure}
\includegraphics{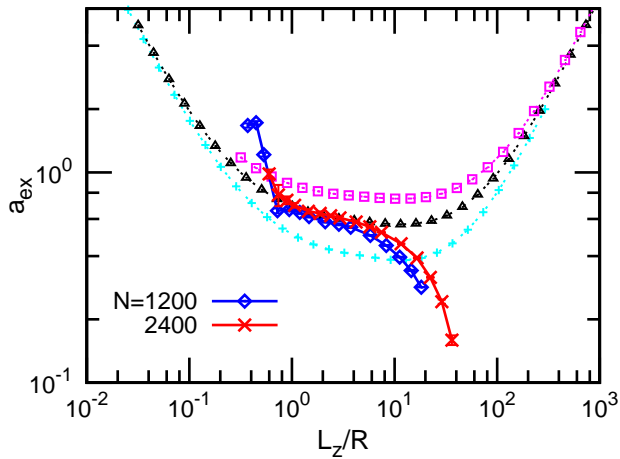}
\caption{\label{fig:arex_tube} (Color online)
Excess area $a_{\rm {ex}}$ dependence on $L_z/R$ 
calculated from 
the simulations and the 
perturbation theory Eq. (\ref{eq:exarea_f}).
For the simulations,  $a_{\rm {ex}}$ is shown
in the range
$ 10 \le L_z/\sigma \le 72$ and $ 18 \le L_z/\sigma\le 144$
for $N=1200$ ($\diamond$)
 and $2400$ ($\times$) at $k_\alpha = 10$, respectively. 
The analytical data from Eq. (\ref{eq:exarea_f})
are also shown
for $L_z/l=10^2$ ($+$), 
$10^3$ ($\triangle$), 
and $10^4$ ($\Box$).
}
\end{figure}

Fournier {\it et al.} \cite{Fournier2007}
derived the dependence of the normalized excess area 
\begin{equation}
a_{\rm {ex}}= \frac{\kappa\Delta A}{k_{\rm B}T A_{\rm p}} = \frac{R}{4\pi L_z} \sum_{m,\overline{q} }
\frac{m^2 +\overline{q}^2}{ (m^2-1)^2 + \overline{q}^2 (\overline{q}^2 + 2m^2) },
\label{eq:exarea_f}
\end{equation}
on the axial tension $\sigma_z$,
from the undulation spectrum
Eqs. (\ref{eq:fluctuation}) and (\ref{eq:wvn_cyl}), 
where $\Delta A = A-A_{\rm p}$.
They predicted that a higher axial tension generates
an increase in the normalized excess area 
owing to the enhanced Goldstone mode fluctuations, 
contrary to the case for planar membranes.
Figure \ref{fig:arex_tube} shows 
the $L_z/R$ dependence of
the excess area $a_{\rm {ex}}$
obtained from the perturbation theory [Eq. (\ref{eq:exarea_f})] and the simulations.
When $a_{\rm {ex}}$ is plotted for $L_z/R$,
the size dependence of  $a_{\rm {ex}}$ from Eq. (\ref{eq:exarea_f})
is seen only for the middle region of $L_z/R \sim 10$,
and then, all the three curves converge at 
$L_z/R \to 0$ and $L_z/R \to \infty$.
The enhanced fluctuations in the azimuth or axial direction
generates a large  $a_{\rm {ex}}$  at $L_z/R \to 0$ or $L_z/R \to \infty$, respectively.
For $L_z/R \lesssim 10$, 
our simulation shows good agreement with their prediction.
However, for $L_z/R \gtrsim 10$, 
$a_{\rm {ex}}$ decreases in the simulation but increases in their theory.
Thus, the thermal undulations of longer tubes
are suppressed in the simulations.
This discrepancy may be caused by the suppression of the protrusion
modes or the effects of the higher-order terms of the perturbations.
Further study is necessary to clarify the origin of this difference.

When the solvent is explicitly taken into account
or when bilayer membranes have a low flip-flop frequency,
the bending rigidity is difficult to simulate using tubular membranes.
The tubular membranes would, in such a case,
exhibit very slow relaxation to the thermal equilibrium state,
considering that a radius variation of the tubular membrane 
accompanies changes in the tube volume
and in the area difference between the two leaflets.
Therefore, the Laplace pressure needs to be taken into account or
an additional numerical technique is required 
to exchange the solvent particles or lipids between
the upper and the lower sides of the bilayers.

\begin{figure}
\includegraphics{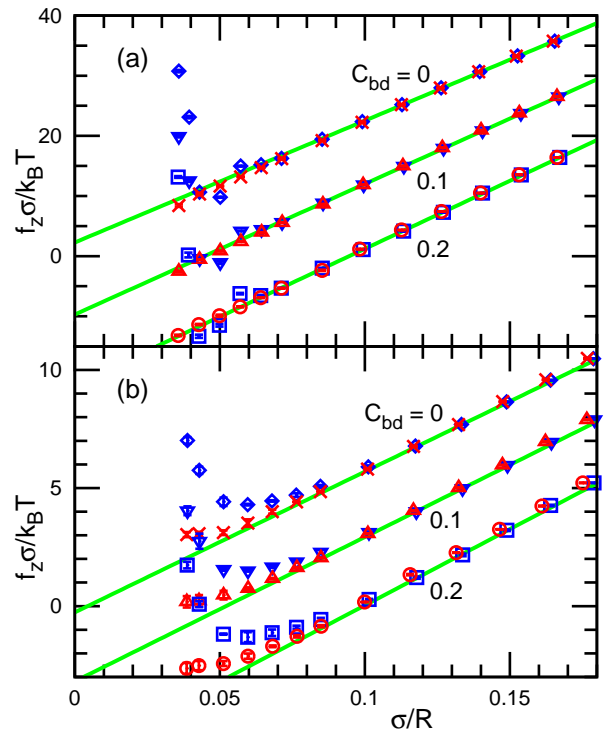}
\caption{\label{fig:fz_spc} (Color online)
Force $f_z$ dependence on the radius $R$ of the membrane tube
for the spin membrane model at $\varepsilon=5$ and $C_{\rm {bd}}=0$, $0.1$, and $0.2$.
(a) $k_{\rm {bend}}=k_{\rm {tilt}}=20$. (b) $k_{\rm {bend}}=k_{\rm {tilt}}=5$.
The symbols ($\diamond, \triangledown, \Box$) and ($\times,\triangle, \circ$) 
represent $N=1200$ and $N=2400$, respectively.
The solid lines are obtained by linear least squares fits.
 }
\end{figure}

\section{measurement of spontaneous curvature} \label{sec:spcv}

\begin{figure}
\includegraphics{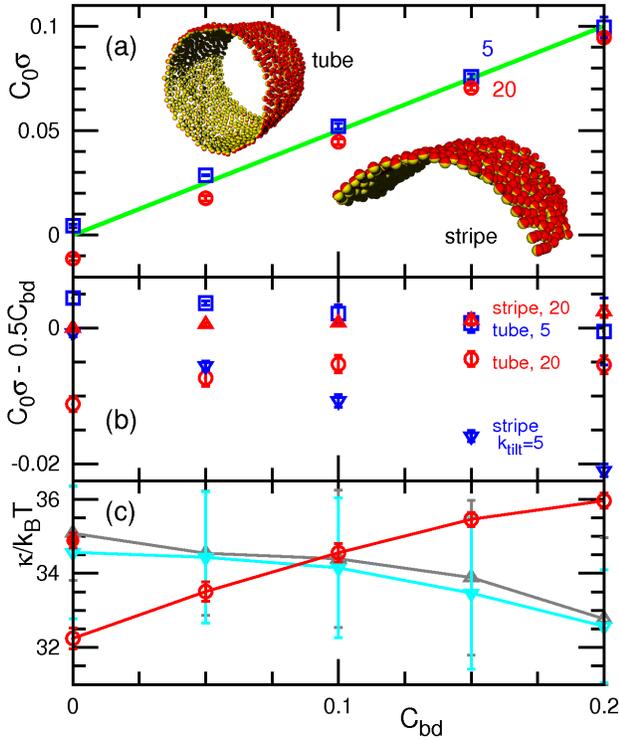}
\caption{\label{fig:c0_spc} (Color online)
Parameter $C_{\rm {bd}}$ dependence of (a), (b) the spontaneous curvature $C_0$
and (c) the bending rigidity $\kappa$ for the spin membrane model at $\varepsilon=5$.
(a) The symbols ($\Box$) and ($\circ$) represent $C_0$ obtained from the membrane tube 
for $k_{\rm {bend}}=k_{\rm {tilt}}=5$ and $k_{\rm {bend}}=k_{\rm {tilt}}=20$, respectively.
The solid line shows the line $C_0 \sigma= 0.5 C_{\rm{bd}}$.
The snapshots of a membrane tube and strip are shown in the inset 
at $N=1200$ and $400$, respectively,
for $L_z=24\sigma$,
$k_{\rm {bend}}=k_{\rm {tilt}}=20$, and $C_{\rm {bd}}=0.2$.
(b) The symbols ($\Box, \circ$) and ($\triangledown, \triangle$) represent the relative
spontaneous curvature $C_0\sigma - 0.5 C_{\rm{bd}}$
obtained from the membrane tubes and membrane strips, respectively.
(c) The opened or closed circles represent the values estimated
by the fits with two fit parameters ($\kappa$, $C_0$) or one parameter ($\kappa$),
respectively. The triangles ($\triangle, \triangledown$) represent $\kappa$ estimated by the undulation analysis 
of the planar membranes with $N=1600$ and $6400$, respectively.
}
\end{figure}

Next, we investigate the estimation method of the spontaneous curvature $C_0$
 using the spin membrane model.
We estimate $C_0$ from the axial force of tubular membranes and the shape of membrane strips.
When the membranes have a nonzero spontaneous curvature,
the membrane tube has the lowest free energy at $1/R=C_0$,
where $f_z$ becomes zero (see Eqs. (\ref{eq:F_tube}) and (\ref{eq:tension})).
Figure \ref{fig:fz_spc} shows that $1/R$--$f_z$ lines move down for 
increasing $C_{\rm {bd}}$.
The finite-size effects discussed in Sec. \ref{sec:nocv} for a small $1/R$ are 
also seen for a finite $C_0$ (See Fig. \ref{fig:kappa_tube}(a)).
The spontaneous curvature $C_0$ and the bending rigidity $\kappa$
were estimated by the linear least squares fit to Eq. (\ref{eq:tension})
for $\sigma/R>0.08$ at $N=2400$.
The estimated $C_0$ increases proportionally with $C_{\rm {bd}}$,
as shown in Fig. \ref{fig:c0_spc}(a): $C_0\sigma=0.5C_{\rm {bd}}$.
A small deviation ($\lesssim 0.01/\sigma$) from the line is
almost independent of $C_{\rm {bd}}$ (see Fig. \ref{fig:c0_spc}(b));
therefore, it is considered a systematic error because
the symmetric membrane at $C_{\rm {bd}}=0$ has $C_0$
precisely equal to zero. 

The bending rigidity $\kappa$ is independent of $C_{\rm {bd}}$.
The results coincide with the $\kappa$ estimated from the thermal undulations
of the planar membranes (see Fig. \ref{fig:c0_spc}(c)).
The two methods have a
 slight dependence on $C_{\rm {bd}}$ with opposite tendencies,
but the dependences are smaller than the error bars.

Alternatively, $C_0$ can be estimated from the shape of a membrane strip.
The membrane is connected by the periodic boundary in one ($x$) direction,
whereas it is open with edges in the other ($y$) direction
(see the snapshot in Fig. \ref{fig:c0_spc}(a)).
Because the membrane can freely bend in the $y$ direction,
the mean curvature should be $C_0$.
The flip of the orientation vector $\bm{u}_i$ of the membrane particles 
is not observed for the investigated parameters.
Thus, the membranes can maintain the value of $C_0$ 
homogeneously even when the membrane has open edges.
We calculated the membrane curvature using the second-order moving
least-squares fit \cite{Nog06PRE} with 
the weight function of the potentials $w_{\rm {cv}}(r)$ for strips 
of $N=400$ with $L_z/\sigma=20$ to $30$.
For a large bending rigidity $\kappa= 34k_{\rm B}T$ ($k_{\rm {tilt}}=20$), 
the resulting $C_0$ follows the relation $C_0\sigma=0.5C_{\rm {bd}}$
better than that of the tube estimation.
However, it seems to underestimate $C_0$ 
for a small bending rigidity $\kappa= 9k_{\rm B}T$ ($k_{\rm {tilt}}=5$)
owing to large particle protrusions.
We confirmed that the resulting values were not sensitive to the shape of the weight in the mls fit.
When a larger radius ($r_{\rm {ga}}=2.5\sigma$ and $r_{\rm {cc}}=5\sigma$) 
is used for the weight $w_{\rm {cv}}(r)$,
the differences from the $C_0$ values calculated with the original weight are less than $5$\%.

The bending elasticity generated by the bending and tilt potentials can be
derived from the continuum theory \cite{Helfrich} as discussed in Ref. \cite{nogu11}.
When the orientation vectors ${\bf u}_{i}$ are equal to the normal
 vectors of the membrane without tilt deformation,
the bending and tilt energies are given by
\begin{eqnarray}
U_{\rm {cv}} &=& \int dA\  \frac{\kappa'_{\rm {bend}}}{2}[
(C_1-C'_0)^2 + (C_2-C'_0)^2] \nonumber \\ \label{eq:cv0} 
& & \hspace{0.9cm} + \frac{\kappa'_{\rm {tilt}}}{2}(C_1^2 +C_2^2) \\ \nonumber
 &=&  \int dA\  \frac{\kappa'_{\rm {bend}}+\kappa'_{\rm {tilt}}}{2}
(C_1+C_2-C_0)^2   \\ \label{eq:cv1} 
& & \hspace{0.9cm} - (\kappa'_{\rm {bend}}+\kappa'_{\rm {tilt}})C_1 C_2 + U_0
\end{eqnarray}
in the continuum limit, where
$C_1$ and $C_2$ are two principal curvatures of
the membrane.
The first and second terms in Eq. (\ref{eq:cv0}) are
the contributions of the bending and tilt potentials, respectively.
The spontaneous curvature of the bending potential is given by $C'_0= C_{\rm {bd}}/\bar{r}_{\rm {nb}}$.
The nearest-neighbor distance $\bar{r}_{\rm {nb}} \simeq 1.15\sigma$ is obtained from the radial distribution function.
By assuming a hexagonal packing of the molecules,
the bending rigidities generated by  the bending and tilt potentials are estimated as 
$\kappa'_{\rm {bend}}/k_{\rm B}T= \sqrt{3} k_{\rm {bend}} w_{\rm {cv}}(\bar{r}_{\rm {nb}})$ and
$\kappa'_{\rm {tilt}}/k_{\rm B}T= \sqrt{3} k_{\rm {tilt}} w_{\rm {cv}}(\bar{r}_{\rm {nb}})/2$, respectively.
The bending rigidity is given by their sum; {\it i.e.}, $\kappa=\kappa'_{\rm {bend}}+\kappa'_{\rm {tilt}}$.
Equation (\ref{eq:cv1}) gives the saddle-splay modulus  $\bar{\kappa}= -\kappa$
and the spontaneous curvature $C_0= \{\kappa'_{\rm {bend}}/(\kappa'_{\rm {bend}}+\kappa'_{\rm {tilt}})\}C_{\rm {bd}}/\bar{r}_{\rm {nb}}$ 
with $U_0 =(\kappa'_{\rm {bend}}+\kappa'_{\rm {tilt}})(1/2+\kappa'_{\rm {tilt}}/\kappa'_{\rm {bend}})C_0^2$.
Thus, $\kappa$ and $C_0$ are estimated as 
\begin{eqnarray} \label{eq:kc_t}
\kappa &=& (k_{\rm {bend}} + 0.5 k_{\rm {tilt}})k_{\rm B}T, \\ \nonumber
C_0 &=& \{k_{\rm {bend}}/1.15\sigma(k_{\rm {bend}}+k_{\rm {tilt}}/2)\}C_{\rm {bd}},
\end{eqnarray}
from $w_{\rm {cv}}(1.15\sigma) = 0.56$.
For $k_{\rm {bend}}=k_{\rm {tilt}}$, $C_0= 0.58 C_{\rm {bd}}$.
This relation explains the simulation results very well.
The $16$\% overestimation of the factor ($0.58$) may be
caused by the assumption of a regular hexagonal structure for the fluid state.

Another method to estimate  the spontaneous curvature $C_0$ in a simulation was
proposed by Markvoort {\it et al.} \cite{mark06}.
They made a sigmoidal shape of the membranes with two domains, 
which have opposite spontaneous curvatures.
Then, $C_0$ was estimated from a comparison of the membrane shape
with the energy minimum curve of the continuum theory \cite{mark06}.
The accuracy of this method is likely to be similar to that of the curved strip
since both the methods use the minimum energy shape of the membranes.

Among these three methods, the $C_0$ estimation for tubular membranes
can be applied even for a small bending rigidity $\kappa \sim 10k_{\rm B}T$.
It would be suitable for other solvent-free models,
where the membrane can freely change the tube volume or 
the area difference between the two leaflets in a bilayer membrane.
For membranes with explicit solvents or with slow flip-flop relaxation,
the other two methods, with the curved membrane strip or sigmoidal membrane,
would be easier to apply.

\section{Summary} \label{sec:summary}

We have investigated numerical methods for measuring the bending rigidity $\kappa$ and
spontaneous curvature $C_0$ of fluid membranes.
For planar membranes, $\kappa$ is estimated from the spectrum of
the thermal undulations.
It is found that estimate values show a large dependence on the 
upper-cutoff frequency  $q_{\rm {cut}}$ for the least-squares fits.
Among the investigated fitting methods,
the inverse power-spectrum fit with the extrapolation to $q_{\rm {cut}} \to 0$
gives an accurate estimation.
For tubular membranes, 
 $\kappa$ is estimated from the stretching force and the spectrum of
the thermal undulations.
The estimated $\kappa$ gives a reasonable agreement with the others
for all of three methods 
as well as for previous methods using the anisotropic surface 
tension of a buckled membrane \cite{nogu11a}
and the thermal undulations of quasi-spherical vesicles \cite{Nog06PRE}.
From a comparison of these methods, it is concluded that
 the inverse power-spectrum fit
with the extrapolation
is the best estimation method for simulations.

The excess area $a_{\rm {ex}}$ of tubular membranes is also investigated.
For short tubes, the calculated $a_{\rm {ex}}$ agrees with that obtained
by Fournier {\it et al.}'s perturbation theory \cite{Fournier2007}.
However, with an increasing tube length,  
$a_{\rm {ex}}$ decreases in the simulation
but increases in their calculation.
This difference may be caused by the finite-size effects in the simulations or
the effects of the higher-order terms of the perturbation theory.

The spontaneous curvature $C_0$ is measured from the axial force of tubular membranes
and the average curvature of bent membrane strips.
Both the methods provide a reasonable estimation.
The methods investigated here are also suitable for other membrane simulation models
from atomistic or coarse-grained molecular models to large-scale meshless models.

\begin{figure}
\includegraphics[width=0.82\linewidth]{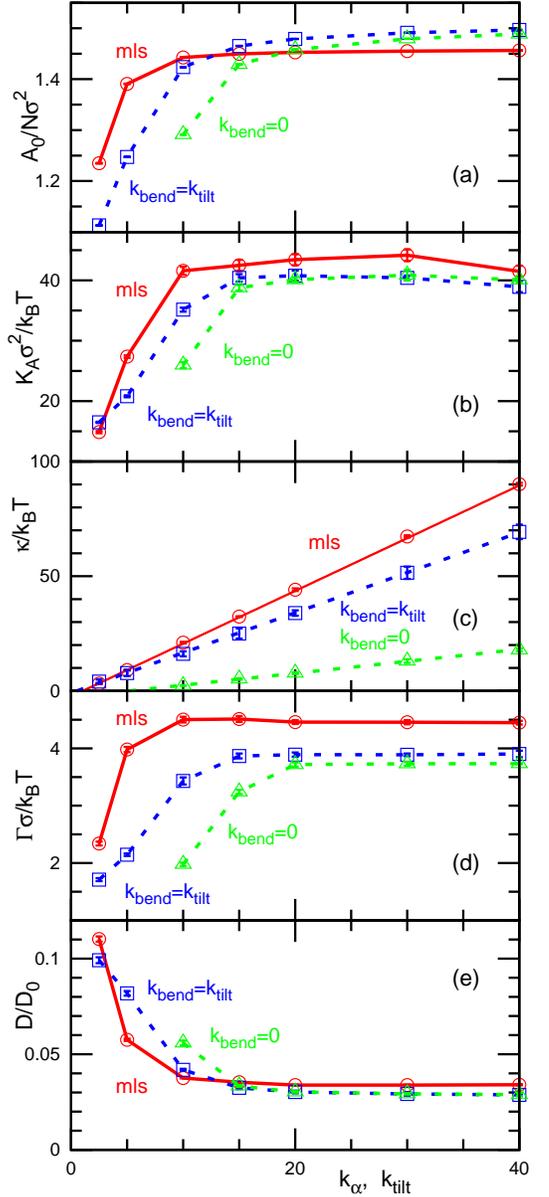}
\caption{\label{fig:mem_k}
(Color online)
Curvature parameter dependence of  (a) the intrinsic area $A_0/N\sigma^2$, 
(b) area compression modulus $K_{\rm A}$,
(c) bending rigidity $\kappa$, 
(d) line tension $\Gamma$, and
(e) diffusion coefficient $D$ for tensionless membranes  at $\varepsilon=4$.
The solid line with circles represents the data for the mls model.
The dashed lines with squares and triangles represent data for the spin model at $C_{\rm {bd}}=0$
for $k_{\rm {bend}}=k_{\rm {tilt}}$ and  $k_{\rm {bend}}=0$, respectively.
The solid and dashed lines in (c) show the linear fits
for $\kappa/k_{\rm B}T=2.3k_{\alpha} -2.4$ and $\kappa/k_{\rm B}T=1.75k_{\rm {tilt}} -0.9$,
as well as for $\kappa/k_{\rm B}T=0.52k_{\rm {tilt}} -2.7$, respectively.
}
\end{figure}

\begin{figure}
\includegraphics[width=0.82\linewidth]{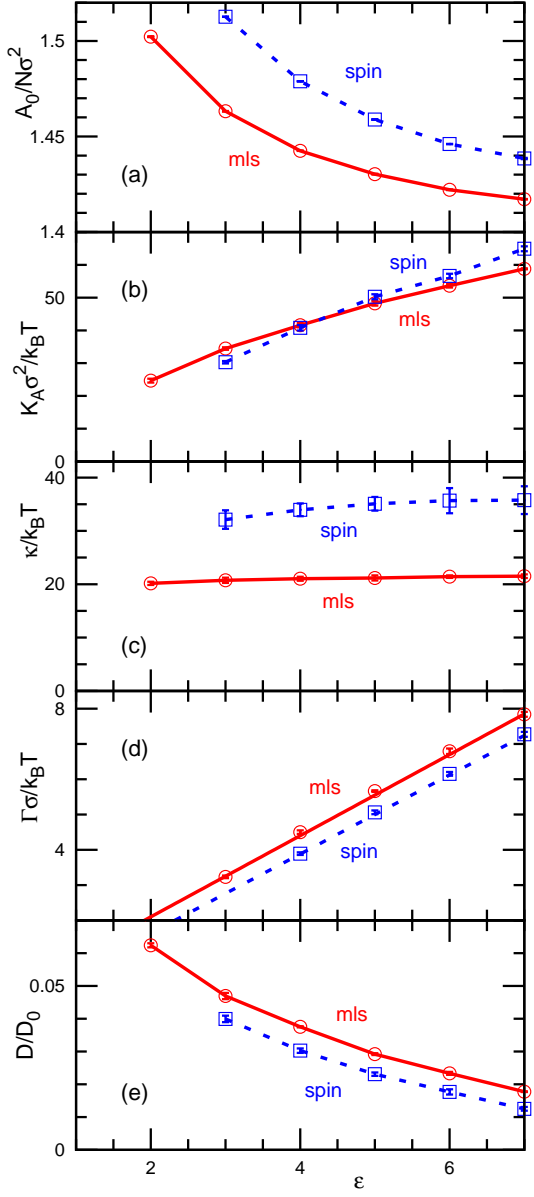}
\caption{\label{fig:mem_e}
(Color online)
Parameter $\varepsilon$ dependence of (a) $A_0/N\sigma^2$,
 (b) $K_{\rm A}$, (c) $\kappa$, (d) $\Gamma$, and
(e) $D$  for the tensionless membranes.
The solid line with circles represents the data for the mls model at $k_{\alpha}=10$.
The dashed line with squares represents the data for the spin model at $k_{\rm {tilt}}=k_{\rm {bend}}=20$
and $C_{\rm {bd}}=0$.
The solid and dashed lines in (d) show the linear fits
$\Gamma \sigma/k_{\rm B}T=1.15\varepsilon -0.2$ and 
$\Gamma \sigma/k_{\rm B}T=1.12\varepsilon -0.6$, respectively.
}
\end{figure}

\begin{acknowledgements}
The authors would like to thank
W. Shinoda, T. Nakamura, T. Taniguchi, H. Wu, M. Deserno, and G. Gompper 
for helpful discussions and comments. 
The numerical calculations were partly
carried out on SGI Altix ICE 8400EX 
at ISSP Supercomputer Center, University of Tokyo. 
This work is supported by KAKENHI (21740308) from
the Ministry of Education, Culture, Sports, Science, and Technology of Japan.
\end{acknowledgements}

\begin{appendix}
\section{membrane properties}

Here, we describe the parameter dependence of the properties
of the tensionless membrane for the mls and spin membrane models.
Figures \ref{fig:mem_k} and  \ref{fig:mem_e} show 
the dependence of five quantities on curvature parameters
($k_{\alpha}$, $k_{\rm {tilt}}$, and $k_{\rm {bend}}$) and attraction strength $\varepsilon$,
respectively.
The membrane is in the fluid phase 
for all ranges of the parameters shown in the figures.

The  intrinsic area $A_0$,
the area compression modulus $K_{\rm A}$, 
the  line tension $\Gamma$ of the membrane edge, and
 the particle diffusion coefficient $D$ 
are almost independent of the curvature parameters when they are sufficiently 
large ($k_{\alpha} \gtrsim 10$ and $k_{\rm {tilt}} \gtrsim 15$).
The bending rigidity $\kappa$ is linearly dependent on the curvature parameters
(see Fig. \ref{fig:mem_k}).
Thus, $\kappa$ can be varied without changing the other membrane properties.
For the spin model, the dependence of $\kappa$ on curvature parameters
can be quasi-quantitatively
explained by Eq. (\ref{eq:kc_t}), derived from the continuum theory.
The slope is only $17$\% or $4$\% higher than the theoretical prediction
for  $k_{\rm {bend}}=k_{\rm {tilt}}$ and  $k_{\rm {bend}}=0$, respectively.
The line tension $\Gamma$ linearly depends on $\varepsilon$,
whereas $\kappa$ is almost independent of  $\varepsilon$ 
(see Fig. \ref{fig:mem_e}).
Thus,  $\kappa$ and $\Gamma$ can be separately varied 
by changing the potential parameters for the spin model as well as for the mls model.

The intrinsic area $A_0$,
the area compression modulus $K_{\rm A}= A_0\partial \gamma/\partial A|_{A=A_0}$,
and the diffusion coefficient $D$ for a tensionless membrane are 
calculated from planar membranes using the method explained in Ref. \cite{Nog06PRE}.
The unit diffusion coefficient is $D_0=\sigma^2/\tau_0$.
The bending rigidity $\kappa$ is estimated using 
the extrapolation method for Eq. (\ref{eq:hq_inv}) at $N=1600$. 
It is $10$\% higher than the values estimated in our previous paper \cite{Nog06PRE},
where  Eq. (\ref{eq:hq_inv}) is fitted with $(q_{\rm {cut}}/\pi)^2=0.05$.

The line tension $\Gamma$ of the membrane edge
is calculated 
from the membrane strips with $N=400$, as follows: \cite{tolp04,reyn08,footnote1}
\begin{equation}
\Gamma =  \langle (P_{yy} + P_{zz})/2 -  P_{xx} \rangle L_{y}L_{z}/2,
\label{eq:lt}
\end{equation} 
since the length of the membrane edge  is $2L_{x}$ and
$\Gamma$ is the energy per unit length of the membrane edge.
The resulting $\Gamma$ values coincide with the values estimated 
from the membrane pore in Ref. \cite{Nog06PRE}.

The difference between  the values is less than $0.1k_{\rm B}T/\sigma$.

\end{appendix}


\begin{thebibliography}{69}
\bibitem[{foo()}]{footnote1}
\bibinfo{note}{A minus sign is missing on the left side of Eq. (17) in Ref.
  \cite{nogu11}.}
\end{thebibliography}

\end{document}